\documentclass[10pt, secnumarabic, amssymb, amsmath, aps, prl, reprint, superscriptaddress]{revtex4-2}

\pdfoutput=1

\usepackage{graphicx}
\usepackage{amsmath}
\usepackage{physics}
\usepackage{color}
\usepackage{natbib}
\usepackage{ulem}

\graphicspath{ {figures/} }

\newcommand{\FX}{\ensuremath{\mathrm{FX}}}

\newcommand{\DX}{\ensuremath{\mathrm{D_A^{*}X}}}
\newcommand{\DBX}{\ensuremath{\mathrm{D_B^{*}X}}}
\newcommand{\DXX}{\ensuremath{\mathrm{D_A^{*}XX}}}
\newcommand{\DBXX}{\ensuremath{\mathrm{D_B^{*}XX}}}

\begin{document}

\title{Correlations between cascaded photons from spatially localized biexcitons in ZnSe}

\author{Robert M. Pettit}
\email{pettitr@umd.edu}
\affiliation{Institute for Research in Electronics and Applied Physics, and Joint Quantum Institute, University of Maryland, College Park, MD 20742}
\affiliation{Intelligence Community Postdoctoral Research Fellowship Program, University of Maryland, College Park, MD 20742}

\author{Aziz Karasahin}
\affiliation{Institute for Research in Electronics and Applied Physics, and Joint Quantum Institute, University of Maryland, College Park, MD 20742}

\author{Nils von den Driesch}
\affiliation{Peter--Gr\"{u}nberg--Institute (PGI-9), Forschungszentrum J\"{u}lich GmbH, 52425 J\"{u}lich, Germany}
\affiliation{JARA--FIT (Fundamentals of Future Information Technology), J\"{u}lich--Aachen Research Alliance, 52062 Aechen, Germany}

\author{Marvin Marco Jansen}
\affiliation{Peter--Gr\"{u}nberg--Institute (PGI-9), Forschungszentrum J\"{u}lich GmbH, 52425 J\"{u}lich, Germany}
\affiliation{JARA--FIT (Fundamentals of Future Information Technology), J\"{u}lich--Aachen Research Alliance, 52062 Aechen, Germany}

\author{Alexander Pawlis}
\affiliation{Peter--Gr\"{u}nberg--Institute (PGI-9), Forschungszentrum J\"{u}lich GmbH, 52425 J\"{u}lich, Germany}
\affiliation{JARA--FIT (Fundamentals of Future Information Technology), J\"{u}lich--Aachen Research Alliance, 52062 Aechen, Germany}

\author{Edo Waks}
\email{edowaks@umd.edu}
\affiliation{Institute for Research in Electronics and Applied Physics, and Joint Quantum Institute, University of Maryland, College Park, MD 20742}

\begin{abstract}
Radiative cascades emit correlated photon pairs, providing a pathway for the generation of entangled photons. The realization of a radiative cascade with impurity atoms in semiconductors, a leading platform for the generation of quantum light, would therefore provide a new avenue for the development of entangled photon pair sources. Here we demonstrate a radiative cascade from the decay of a biexciton at an impurity-atom complex in a ZnSe quantum well. The emitted photons show clear temporal correlations revealing the time--ordering of the cascade. Our result establishes impurity atoms in ZnSe as a potential platform for photonic quantum technologies using radiative cascades.
\end{abstract}

\date{Compiled \today}

\maketitle

Radiative cascades in semiconductors are a promising tool for generating nonclassical and entangled states of light. These cascades are often composed of an exciton-biexciton emission pair, where the biexciton consists of two excitons that form a lower energy complex due to Coulombic interactions \cite{bensonRegulatedEntangledPhotons2000}. The decay of the first exciton initiates a radiative cascade in which the photon emitted by the second exciton may be entangled in polarization \cite{liuSolidstateSourceStrongly2019, wangOnDemandSemiconductorSource2019}, time \cite{jayakumarTimebinEntangledPhotons2014}, or both \cite{prilmullerHyperentanglementPhotonsEmitted2018}. The photon pairs generated through biexciton decay could be used as sources of entangled light with applications in quantum communication \cite{simonQuantumRepeatersPhoton2007, bassobassetQuantumKeyDistribution}, photonic quantum computing \cite{slussarenkoPhotonicQuantumInformation2019}, and quantum metrology \cite{giovannettiAdvancesQuantumMetrology2011}.

A number of material platforms have demonstrated cascaded emission to--date, with the majority of recent work focused on epitaxial quantum dots  \cite{huberSemiconductorQuantumDots2018}. These emitters can support bright exciton-biexciton emission pairs which emit entangled photons on--demand with high radiative efficiency \cite{liuSolidstateSourceStrongly2019, wangOnDemandSemiconductorSource2019}. Single photon emission in a monolayer of WSe2 has also produced radiative cascades in an atomically thin form-factor \cite{heCascadedEmissionSingle2016}. Impurity atoms in semiconductors are another material platform that shows promise for realizing both sources of quantum light and spin--photon interfaces \cite{awschalomQuantumTechnologiesOptically2018}. Although evidence for biexciton formation has been observed in these materials at isoelectronic impurity dyads \cite{marcetChargedExcitonsBiexcitons2010} and other impurity centers \cite{merzExcitonicMoleculeBound1969a, sartiMultiexcitonComplexExtrinsic2013, dottiGermaniumbasedQuantumEmitters2015}, cascaded emission has yet to be demonstrated.

In this letter, we demonstrate a radiative cascade of single photons from an atomic impurity center in a ZnSe quantum well. The quantum well is delta-doped with chlorine donor atoms that act as bright impurity-bound exciton emitters \cite{karasahinSingleQuantumEmitters}. From photoluminescence measurements we observe clear emission line-pairs that exhibit a power dependence consistent with an exciton-biexciton cascade. Time-resolved lifetime and photon correlation measurements conclusively reveal the time--ordered nature of the cascade, and polarization resolved spectroscopy reveals a fine structure splitting of $\sim 290~\mathrm{\mu eV}$ for each pair. The observation of cascaded emission at impurity atoms in ZnSe opens up new opportunities to engineer quantum light emission, particularly in ZnSe, which is emerging as a promising host for single photon emitters \cite{sanakaIndistinguishablePhotonsIndependent2009, sanakaEntanglingSinglePhotons2012, karasahinSingleQuantumEmitters} and potential spin--based qubits \cite{degrevePhotonAntibunchingMagnetospectroscopy2010, sleiterOpticalPumpingSingle2013, ethier-majcherCompleteQuantumControl2014, st-jeanHighFidelityUltrafastInitialization2016, kirsteinExtendedSpinCoherence2021}.

\begin{figure}
\centering
\includegraphics[width=8.6cm]{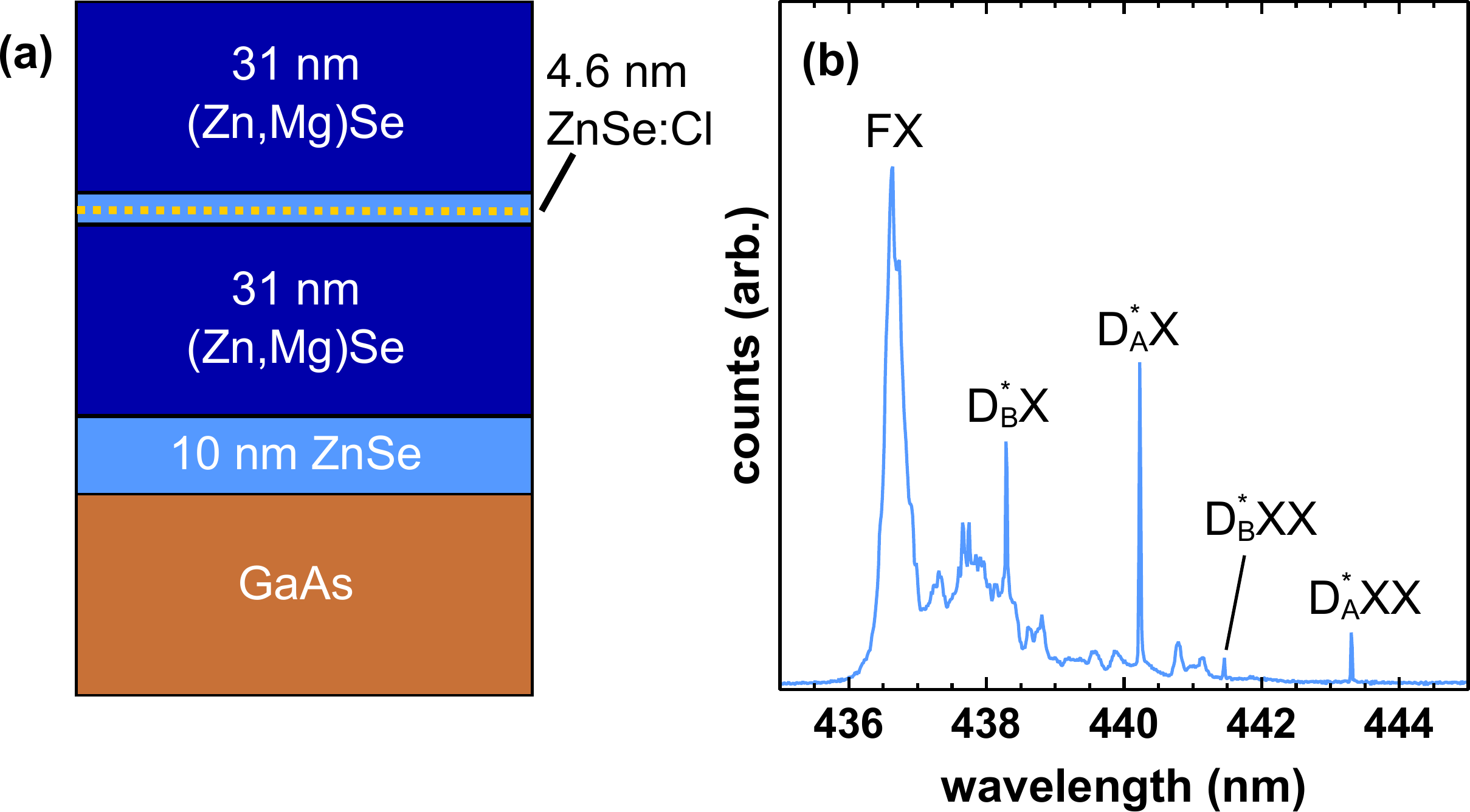}
\caption{Sample and spectral identification of localized emission. (a) Overview of the quantum well structure. (b) Micro--photoluminescence spectrum identifying the quantum well free exciton ({\FX}), along with pairs of localized emission lines {\DX} and {\DXX}, as well as {\DBX} and {\DBXX}.}
\label{fig1}
\end{figure}

Figure \ref{fig1}a illustrates the layer structure of the sample used in this study. The sample consists of an epitaxially grown 4.6~nm ZnSe quantum well delta--doped with chlorine impurity atoms during growth with a density of $\sim 10^{9}$~cm$^{-2}$. The well was capped by 31~nm barrier layers of (Zn,Mg)Se with a Mg concentration of 12\% to provide enhanced carrier confinement and prevent carrier leakage into the GaAs substrate. A 10~nm buffer layer of ZnSe was grown on the GaAs substrate to improve the III-V/II-VI interface. Chlorine was chosen as a dopant because it is an electron donor when replacing selenium in the ZnSe lattice, and has recently been shown to exhibit single photon emission with potential for use as a spin--photon interface \cite{karasahinSingleQuantumEmitters}. We performed measurements of the sample in a helium exchange gas cryostat and used a home--built confocal microscope for optical characterization. The experimental apparatus is discussed in further detail in Supplementary Material section 1. 

Figure \ref{fig1}b shows a photoluminescence spectrum from the sample at 3.6~K, recorded with continuous wave optical excitation above the barrier bandgap at 3.06~eV (405~nm). We observe emission from the free heavy hole exciton (\FX), in addition to a discrete pair of lines {\DX} and {\DXX} at lower energy which are the focus of this letter. We also observe a second pair of lines visible in the spectrum, {\DBX} and {\DBXX}, which we attribute to emission from a separate impurity center. We present a characterization of this second pair of lines in Supplementary Material section 2.

\begin{figure}[b]
\centering
\includegraphics[width=8.6cm]{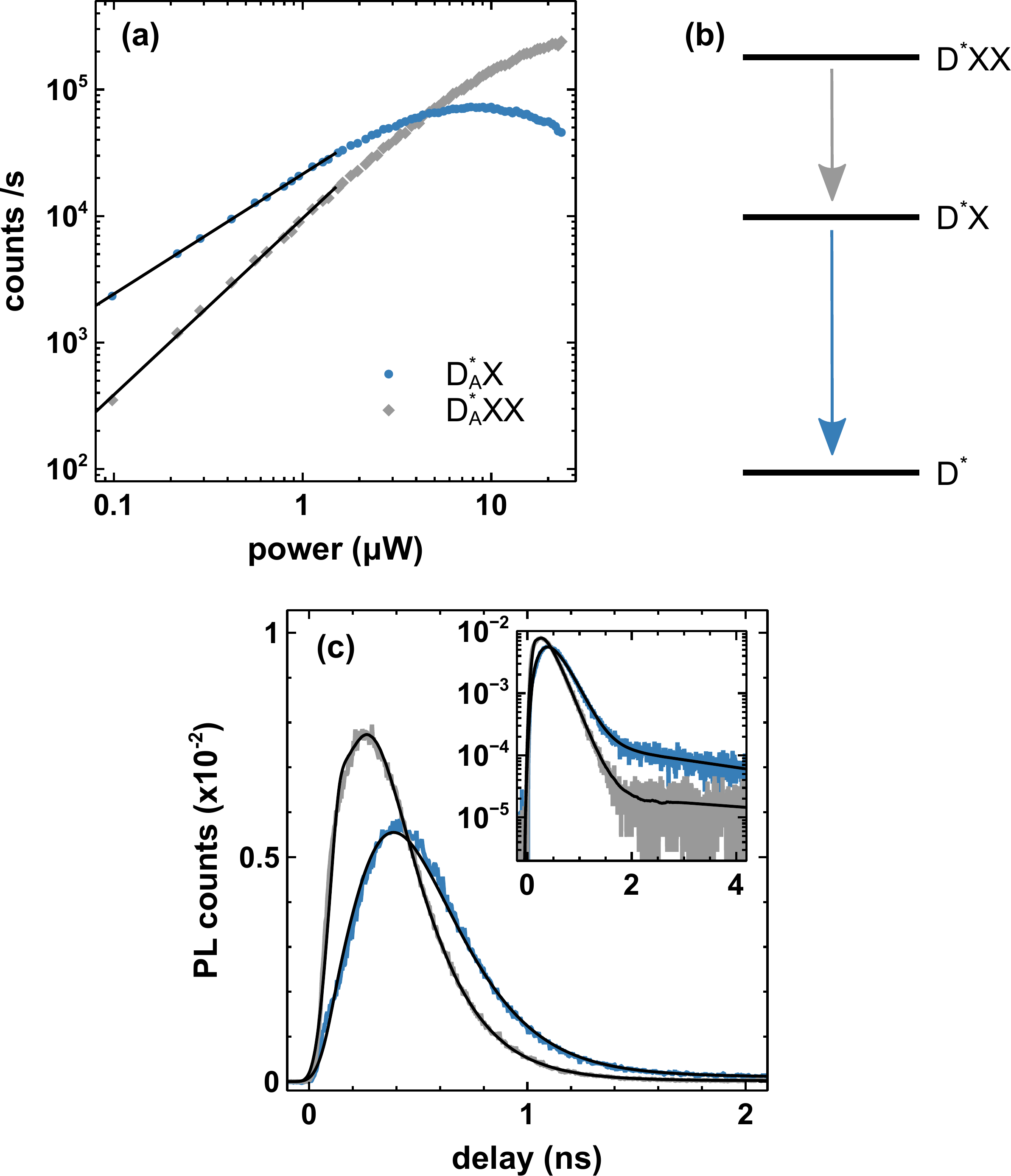}
\caption{Exciton and biexciton classification. (a) Excitation power dependence of the localized emission lines {\DX} and {\DXX}. (b) Energy level structure showing the expected cascaded relationship between the localized biexciton and exciton complexes. (c) Time--resolved photoluminescence collected from the {\DX} and {\DXX} lines. The inset presents the data on a logarithmic scale to better highlight the slower time constants.}
\label{fig2}
\end{figure}

Figure \ref{fig2}a shows the excitation power dependence of the emission intensity of the {\DX} and {\DXX} lines. We observe markedly different behaviors for each line under increasing excitation power. In the low excitation power regime we fit the intensity of each line to $I(p)\propto p^k$, where $k$ is a fit parameter. For the {\DX} line we observe the exponent $k=0.95$, indicating nearly linear dependence of the emission intensity on the excitation power. On the other hand, for the {\DXX} line we observe $k=1.40$, indicating a super--linear dependence on the excitation power. The nearly linear power dependence of the {\DX} line and the superlinear power dependence of the {\DXX} line, followed by saturation of the emission intensity at higher excitation powers, are characteristic of localized exciton and biexciton emission pairs \cite{ulrichTriggeredPolarizationcorrelatedPhoton2003}. 

To demonstrate that emission from {\DX} and {\DXX} are cascaded, as shown by the expected energy level diagram in Fig. \ref{fig2}b, we perform time--resolved lifetime measurements of each line in the emission pair. We excite the sample with 3~ps optical pulses with a photon energy of 3.06~eV from a frequency doubled Ti:Sapphire laser and filter the emission with a grating monochromator. Figure \ref{fig2}c shows the time--resolved measurements of the optical emission from the {\DX} and {\DXX} lines. In all panels the recorded transients are normalized by the total number of recorded counts to better show the relative shape of each decay. The solid lines are fits to the data using a multiexponential function convolved with the detector response, shown in Supplementary Material section 3. Emission from the {\DXX} line decays with two timescales. We observe a fast time constant of $\tau^{\mathrm{XX}}_{1}=106(2)$~ps first and a small contribution from a slower time constant of $\tau^{\mathrm{XX}}_{2}=4.6(7)$~ns second, where the value in parenthesis gives the uncertainty of the last digit. Emission from the {\DX} line rises more slowly and reaches its peak intensity only after the onset of {\DXX} decay. The delay between peak emission of the {\DXX} and {\DX} lines is $\sim 110$~ps, comparable to $\tau^{\mathrm{XX}}_{1}$, suggesting that the decay of {\DXX} feeds into the {\DX} state. The decay of the {\DX} line also exhibits a biexponenial decay with a fast time constant of $\tau^{\mathrm{X}}_{1}=142(3)$~ps and a slower time constant of $\tau^{\mathrm{X}}_{2}=3.5(1)$~ns. The slow decay components, $\tau_2$, present in both {\DX} and {\DXX} emission may indicate the influence of deeper trapping states, dark excitons, or possible carrier re--capture from the quantum well.

\begin{figure*}
\centering
\includegraphics[width=16cm]{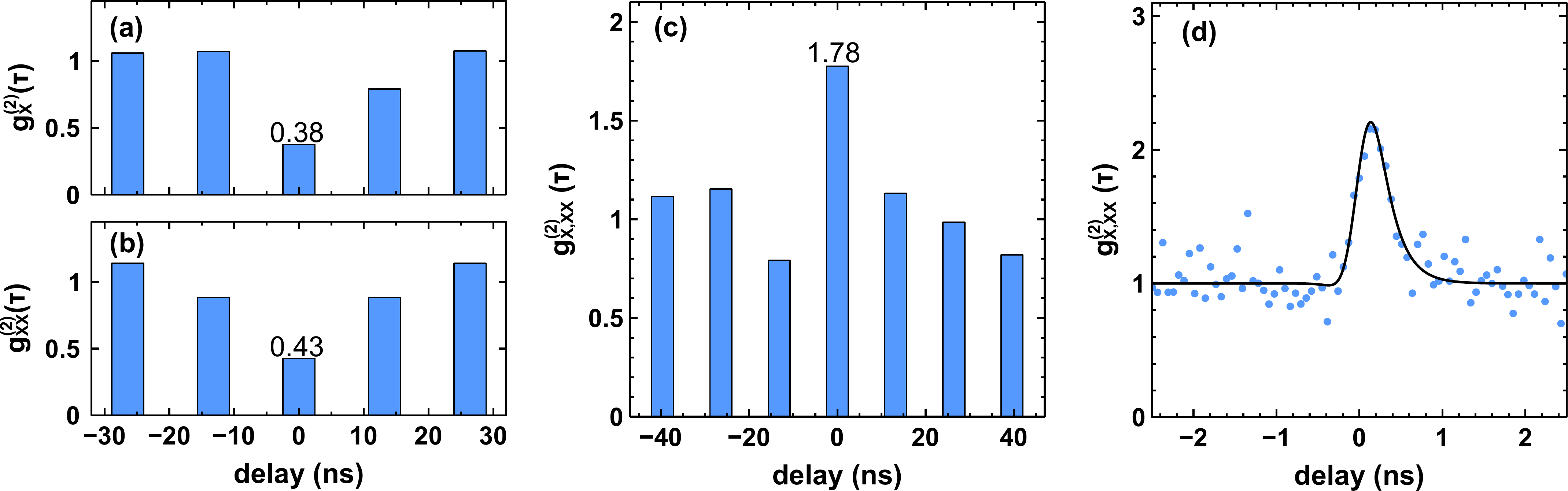}
\caption{Single photon correlation measurements. (a,b) Autocorrelation measurements of (a) the localized exciton line, $g^{(2)}_{X}(\tau)$, and (b) the localized biexciton line, $g^{(2)}_{XX}(\tau)$. (c,d) Cross--correlation measurements between the localized exciton and biexciton lines,  $g^{(2)}_{X, XX}(\tau)$,  with (c) pulsed excitation and (d) continuous wave excitation above the quantum well barrier.}
\label{fig3}
\end{figure*}

To more definitively demonstrate the cascaded nature of the emission from {\DX} and {\DXX} we perform second--order photon correlation measurements. Figures \ref{fig3}a and \ref{fig3}b show the second--order autocorrelations of the {\DX} and {\DXX} lines respectively,  recorded under pulsed excitation using a Hanbury--Brown and Twiss setup. Each correlation is anti--bunched, given by $g^{(2)}(0) < 0.5$, which verifies the single photon nature of the emitted light. For {\DX} we measured $g^{(2)}_{X}(0) =0.38$, while for {\DXX} we measured $g^{(2)}_{XX}(0) =0.43$. In order to verify the correlation between the cascaded single photons, we perform cross--correlation measurements between the {\DXX} and {\DX} emission lines. For the cross--correlation measurements the emission from the sample was split with a 50:50 beam splitter and the {\DX} and {\DXX} lines were filtered through separate monochromator slits. Figure \ref{fig3}c shows the cross--correlation with pulsed excitation which displays significant photon bunching of $g^{(2)}_{X,XX}(0) = 1.78$. The observation of photon bunching, given by $g^{(2)}_{X,XX}(0) >1$, establishes the enhanced probability of detecting a single photon from the {\DX} line after first detecting a single photon from the {\DXX} line. Photon bunching provides conclusive evidence for the cascaded relationship between the two emission lines and is therefore direct evidence for the presence of the localized biexciton initiating a radiative cascade.

Figure \ref{fig3}d shows the measured cross--correlation using continuous wave excitation instead of pulsed excitation. This measurement provides further insight into the radiative cascade by revealing the characteristic temporal asymmetry of $g^{(2)}_{X,XX}(\tau)$ expected near $\tau=0$ for cascaded emission \cite{moreauQuantumCascadePhotons2001}. The asymmetry is a signature of $g^{(2)}_{X,XX}(\tau)$ sharply transitioning between antibunching for $\tau<0$ and bunching for $\tau>0$. Antibunching arises in the continuous wave measurement because detection of a photon from the {\DX} state projects the system into the impurity ground state, out of which a biexciton photon cannot be emitted. The temporal response of our correlation experiment ($\sim 145$~ps) prevented the direct observation of antibunching for $\tau<0$ at the required time scale, but still allowed the observation of the resulting asymmetry of the bunching profile. The asymmetry is revealed by the sharp rise of the bunching profile for $\tau<0$, while the profile decays with a time $\sim \tau^{\mathrm{X}}_{1}$ for $\tau>0$. The fit function is a solution to a three--level rate equation model taking into account the experimentally measured lifetimes of the {\DX} and {\DXX} states convolved with the measurement system response \cite{moreauQuantumCascadePhotons2001}.

\begin{figure}
\centering
\includegraphics[width=8.6cm]{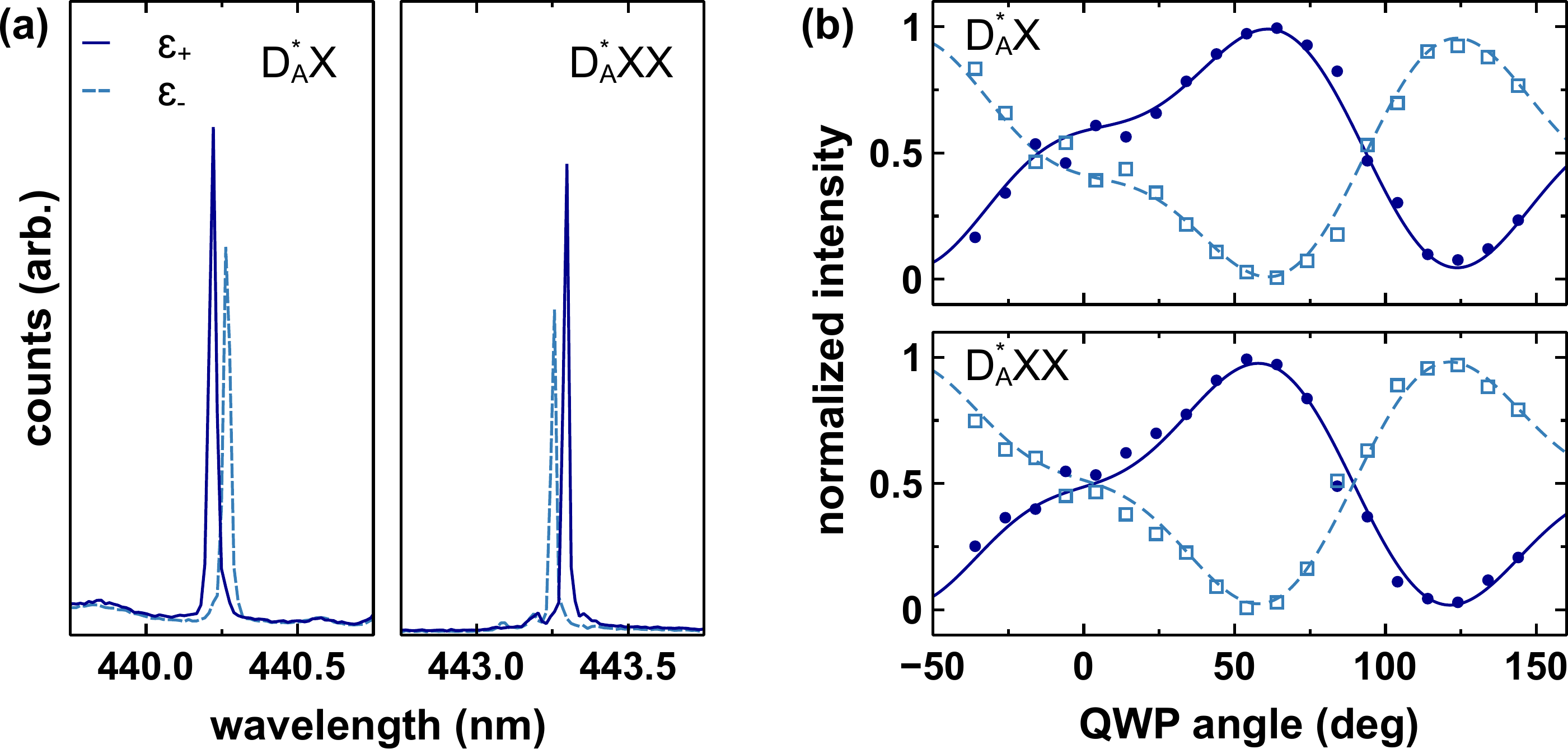}
\caption{Polarization resolved spectroscopy. (a) Measured polarized fine structure of the {\DX} and {\DXX} lines. (b) Polarization analysis of the {\DX} and {\DXX} fine structure as a funciton of quarter wave plate angle.}
\label{fig4}
\end{figure}

Finally, we examine the polarization of the {\DX} and {\DXX} lines in order to study the fine structure of the emission complex. We find each to be a polarized doublet composed of orthogonal elliptical polarizations. Figure \ref{fig4}a shows polarization resolved spectra of each line.  The spectra reveal a fine structure splitting between the orthogonal elliptical components of $\sim \pm 290~\mathrm{\mu eV}$ for the {\DX} and {\DXX} lines, respectively. The equal magnitude but opposite sign of the fine structure splitting suggests that the splitting originates in the localized exciton state {\DX} \cite{kulakovskiiFineStructureBiexciton1999}.

To quantify the ellipticity of the emission from each fine structure component, we performed polarization analysis using a combination of a quarter--wave plate, half--wave plate, and linear polarizer. Rotating the quarter--wave plate in the analyzer modulated the intensity of each line as shown in Fig. \ref{fig4}b. We characterize the ellipticity of each line based on the fits in Fig. \ref{fig4}b. From each fit we determined the ellipticity angle $\chi$, defined in terms of the polarization ellipse as $\tan \chi = \mathrm{\varepsilon}$, where $\mathrm{\varepsilon}$ is the ratio between the ellipse's minor and major axes. The angle $\chi$ therefore assumes the value $\chi=0$ ($\pm\pi/4$) for linear ($\sigma_{\pm}$ circular) polarizations \cite{BasicPropertiesElectromagnetic1999}. A description of the characterization method is provided in Supplementary Material section 4. For the {\DX} doublet we found $\chi=\pm 0.13(1)\pi$, while for the {\DXX} doublet we found $\chi=\pm 0.14(1)\pi$ in close similarity. 

The ellipticity of the fine structure components observed here contrasts with typical observations of neutral biexciton cascades in self--assembled quantum dots where the fine structure components are well described by orthogonal linear polarizations \cite{liuSolidstateSourceStrongly2019, wangOnDemandSemiconductorSource2019}. The magnitude of the fine structure splitting and optical polarization of each line are closely related to the confinement symmetry and charge configuration of the localized excitons, and typically emerges from a confinement symmetry below $D_{2d}$ \cite{kulakovskiiFineStructureBiexciton1999}. Possible candidates for the observed impurity center are a neutral chlorine donor ($D^0$), an ionized Chlorine donor ($D^+$), which is stable in ZnSe and known to have a higher excitonic binding energy than the neutral donor \cite{deanDonorBoundexcitonExcited1981}, or potential multi--donor complexes which cannot be ruled out. In particular, the observed elliptical polarization of the fine structure components may be explained by anisotropic electron--hole exchange in the presence of a highly charged biexciton cascade \cite{akimovFineStructureTrion2002,changOpticalTransitionsPositively2009}. Further understanding of the impurity center's charge configuration will enable strategies for minimizing the observed fine structure splitting, which is a requirement for realizing a source of entangled photons. Strategies for minimizing the fine structure splitting may include tuning with electric or magnetic fields or strain \cite{bennettElectricfieldinducedCoherentCoupling2010, stevensonMagneticfieldinducedReductionExciton2006, seidlEffectUniaxialStress2006}. 

In conclusion, we have demonstrated cascaded single photon emission from a radiative exciton--biexciton cascade in a chlorine doped ZnSe quantum well. Excitons and biexcitons were observed to form with linear and superlinear excitation power dependence, respectively. The resulting radiative cascade was clearly observed in both time--resolved photoluminescence and single--photon correlation measurements. The observed fast decay times of 106 ps and 142 ps for the biexciton and exciton further highlight the promise of this system to produce fast and bright emission of single photons and photon pairs. These results provide a path for the development of on--demand sources of entangled photons with impurity atoms in semiconductors. 

\begin{acknowledgments}
This work is supported by the Air Force Office of Scientific Research (grant No. FA95502010250), The Maryland--Army Research Lab Quantum Partnership (grant No. W911NF1920181), and Deutsche Forschungsgemeinschaft (DFG, German Research Foundation) under Germany's Excellence Strategy--Cluster of Excellence Matter and Light for Quantum Computing (ML4Q) (EXC 2004/1 -- 390534769). R.M.P. acknowledges support through an appointment to the Intelligence Community Postdoctoral Research Fellowship Program and the University of Maryland, administered by Oak Ridge Institute for Science and Education through an interagency agreement between the U.S. Department of Energy and the Office of the Director of National Intelligence.
\end{acknowledgments}


%

\end{document}